# Effect of salt concentration on the solubility, ion-dynamics, and transport properties of dissolved vanadium ions in lithium-ion battery electrolytes: Generalized solubility limit approach (Part II)

*Arijit Mitra[a], Saptarshi Das[b], Debasish Das[c], Subhasish B. Majumder[b], Siddhartha Das[a*]*

[a] Department of Metallurgical and Materials Engineering, Indian Institute of Technology Kharagpur, Kharagpur, West Bengal, India – 721302

[b] Materials Science Center, Indian Institute of Technology Kharagpur, Kharagpur, West Bengal, India – 721302

[c] School of Nano Science and Technology, Indian Institute of Technology Kharagpur, Kharagpur, West Bengal, India – 721302

**Corresponding Author**

  * Email: sdas@metal.iitkgp.ernet.in; Phone - +91-3222-283256

**ABSTRACT**

In this article, we study the transport properties of superconcentrated electrolytes using Molecular Dynamics simulations, which have been shown experimentally to retard elemental dissolution in vanadium containing cathode materials. Five compositions between one and seven molar lithium bis(trifluoromethanesulfonyl)imide in 1,3-Dioxolane and 1,2-Dimethoxyethane solvent mixture are studied using non-polarizable Optimized Potentials for Liquid Simulations - All Atom force field. The simulated physico-chemical properties such as ionic conductivity, self-diffusion coefficients, and density are observed to match well with the results obtained through experiments. Radial Distribution Function analysis reveals a strong co-ordination between salt anions and vanadium cations as the electrolyte transitions from a salt-in-solvent type to solvent-in-salt type electrolyte. A high anion content in the first solvation shell of vanadium cations is observed for solvent-in-salt type electrolytes, through ion-clustering calculations. Solvation free energy calculations using Free Energy Perturbation method indicate that the active material dissolution should be retarded by using superconcentrated electrolytes. Ion-dynamics of the clusters reveal that vanadium cation transport occurs against its concentration gradient due to strong coulombic interactions with the salt anions in superconcentrated electrolytes. The improvement in the cycleability of several vanadium containing cathode materials provides a robust proof for the theoretical framework described in this manuscript.

## 1. INTRODUCTION

With the world moving ahead with the adoption of electric vehicles and other carbon emission free technologies, a necessity of creating batteries with improved energy and power density has suddenly emerged in the research community in order to realize these goals. Several non-traditional lithium-ion battery chemistries are being explored, in parallel to the innovations being made in the fabrication aspects of the lithium-ion batteries. Vanadium containing cathode materials are considered to be attractive for lithium-ion batteries due to the high specific capacity offered by them.[1-7] These materials are still attractive to the research community in spite of the lower working potential range offered by them, since the specific energy density is observed to be better than the ones offered by conventional LFP, LMO, NMC chemistries. However, a major cause of concern in such materials has been their poor cycling performance, which ultimately limits their use in lithium-ion batteries. Over the past few decades, several strategies have been developed and experimentally tested to improve the cycleability of these vanadium containing cathode materials. Most of these strategies are inclined towards engineering the material itself to prevent continuous irreversible phase transitions within them.[8-11] Such strategies include oxide coating over the particles, site doping, particle shape and morphology modifications, etc.[4, 12-15] These strategies demonstrate improvements in the cycleability of these materials. However, they still cannot compete with the metrics offered by other commercial cathode materials in terms of cycleability, and hence, are unsuitable for commercialization.

One of the major issues which are observed in vanadium containing cathode materials is active material dissolution during battery operation.[16-19] This issue is quite neglected in the literature even though it is directly related to the cycling performance of these materials. Active material dissolution can not only cause a loss of active material, but can also foul both the electrolyte and the anode, thereby degrading their performance as well.[20, 21] A handful of publications in the literature report and address this issue which is prevalent in these materials.[16] Since the dissolution products ultimately enter the electrolyte in the batteries, engineering them (instead of the cathode material) in such a way that they reject the dissolution products can result in significant improvement of the cycleability. Recently, Mitra et al. demonstrated the use of highly concentrated electrolytes to improve the cycleability of a high specific capacity Kazakhstanite phase ($Fe_5V_{15}O_{39}(OH)_9.9H_2O$) via the solubility limit approach.[unpublished work] The material undergoes severe active material dissolution in conventional Li-ion electrolytes, which is retarded by switching to a superconcentrated electrolyte. They argued and demonstrated experimentally that by populating the electrolyte with ions instead of solvent molecules, a solubility limit can be created which prevents the dissolution of the Kazakhstanite phase. They demonstrated through theoretical derivations and experimental validations that the improved cycleability can partly be attributed to the improvement in the electrochemical characteristics of the SEI and the passivation layer which forms over the metal anode surface. Cambaz et al. also demonstrated similar results, where they reported an improved cycleability of cation-disordered $Li_{2-x}VO_2F$ using concentrated electrolytes by suppressing vanadium dissolution.[16] However, a concrete framework on the mechanism behind this suppression is not provided by the authors. Fairly recently, Dubouis et al. studied the insertion electrochemistry in soluble layered halides with the superconcentrated electrolytes, and provided a framework similar to the one proposed by Mitra et al.[22] They argued and proved that the positive attribute of superconcentrated electrolytes against the solubility of inorganic compounds is rooted in a thermodynamic rather than a kinetic effect.[22] However, it is incorrect to ignore the kinetic effects in such systems wherein ion-ion coordination is predominant. A deeper understanding of the behavior of such electrolytes is needed in order to confirm whether dissolution is itself arrested or not. Also, verification is needed on its extension as a generalized strategy to improve the cycleability of any vanadium containing cathode materials.

In this manuscript, the answers to the above posed problems are discussed in detail through Molecular Dynamics simulations and suitable experimentations. The simulations are focused towards understanding the changes in the behavior of the vanadium ions as the electrolyte is transitioned from a salt-in-solvent type to a superconcentrated solvent-in-salt type. The simulated values of few macroscopic observables such as ionic conductivity, density, viscosity, and self-diffusion coefficients are compared with the experimental ones performed by us as well as those reported in the literature to prove the validity of the model. Finally, an experimental validation of the conclusions drawn from the simulation results is provided to support the theory discussed. This work is a continuation to results reported earlier by Mitra et al., wherein the solubility limit approach is introduced and demonstrated on the Kazakhstanite phase.[unpublished work] The results described in this manuscript indicate that the use of superconcentrated electrolytes should improve the long-term cycleability of nearly every vanadium containing cathode materials, which have an inherent issue of active material dissolution.

## 2. EXPERIMENTAL SECTION

### 2.1.Molecular Dynamics Simulations Methodology

A stable version of Large-scale Atomic/Molecular Massively Parallel Simulator (LAMMPS – August 2019) was used for all the simulations.[23, 24] All the simulations were run on the Param-Shakti HPC Cluster, set up at IIT Kharagpur. Additional codes for cluster analysis were written in Python.

A three dimensional cubic simulation box consisting of Vanadium, Lithium, Hydroxide and Bistriflimide ions, along with 1-3 Dioxolane and 1-2 Dimethoxyethane molecules, was constructed using Packmol.[25] The structures of Vanadium, Lithium, Hydroxide and Bistriflimide ions, along with 1-3 Dioxolane and 1-2 Dimethoxyethane molecules, were constructed using Avogadro.[26] Depending on the concentration of the salt added to solvent, the number of these entities is varied in the initial simulation box, which is tabulated in Table 1. All the initial configurations were relaxed using the energy minimization criteria, which adjusts the atom co-ordinates till a convergence criterion of 1E-05 was attained. The equilibration runs were first performed under NVE integration coupled with Langevin thermostat maintaining the temperature at 323K, to stabilize the phase space trajectory. This step was followed up by further runs in the NPT ensemble (323K temperature and 1 atm pressure) to allow for the simulation box to attain a stable volume and lowest possible energy configuration. Finally, the temperature was lowered to 300K in the NPT ensemble at 1atm pressure to generate equilibrated systems for further production runs. The total time for equilibration was over 20 ns to ensure that proper mixing has taken place, and any stray effects from the initially assembled systems are eliminated.

The bonded and non-bonded force-field parameters for DOL and DME were taken from OPLS-AA, which is also available within the Moltemplate package.[27, 28] The non-bonded interactions consist of pairwise Lennard-Jones and coulombic interactions, with a cutoff distance of 10Å. The long range electrostatic interactions were computed using the PPPM method, with an accuracy of 0.0001. The LJ pairwise interaction parameters of unlike atoms were calculated from the Lorentz-Berthelot mixing rules. The bonded and non-bonded force-field parameters for TFSI anion was taken from the results reported by Lopes et al.[29] The bonded and non-bonded force-field parameters for hydroxide ion were taken from the TIP3P water model.[30] The force-field parameters for lithium and vanadium ions were taken from results reported by Pluhařová et al. and Gupta et al., respectively.[31, 32] The partial charges for DOL, DME, and TFSI were obtained from Atomic Charge Calculator II, maintained by Masaryk University.[33] The partial charges were scaled down by a factor of 0.67 ($\sim \sqrt{1/\eta_\infty}$), to reflect the charge screening effect absent in non-polarizable force-fields.[34, 35]

The density of the electrolytes was calculated by taking the average density obtained post equilibration runs. The self-diffusion coefficients of the ions were calculated from the Einstein form of the Green-Kubo relations, where the intercept of the Mean Square Displacement (MSD) vs time in log-log scale is equal to the self-diffusion coefficient. The ionic conductivity of the electrolytes was calculated from the Einstein form of the ionic conductivity, while accounting for ionic interactions, as described by France-Lanord et al.[36] The production runs for calculation of MSD and ionic conductivity was of duration of 20 ns and 5 ns, respectively. The shear viscosity of the electrolytes was calculated from the Reversible Equilibrium Molecular Dynamics (REMD) scheme described by Muller-Planthe.[37] The de-solvation energy of the vanadium ions in the electrolytes was calculated from the Free-Energy Perturbation (FEP) and Finite Difference Thermodynamic Integration (FDTI) technique, wherein the non-bonded LJ parameters and the partial charges of V-OH assembly was varied while measuring the total energy change in the system.[38, 39] The partial charges and the LJ parameters were ramped down from a scale factor of 1 to 0 in a step of 0.05, while equilibrating the system for 500 ps before the next scaling was performed. Ion dynamics was studied by implementing self-designed algorithms on the output generated after MD runs.

### 2.2. Materials Synthesis and Structural Characterization of the raw materials

Vanadium Pentoxide (>=98%) was purchased from Sigma Aldrich, and used in its as-received condition. $LiV_3O_8$ was synthesized as per the recipe reported by Pan et al.[40] Commercial $LiMn_2O_4$ was purchased from Gelon LIB group, and used in its as-received form. The phase purity of the as-received $V_2O_5$, as-prepared $LiV_3O_8$, and as-received $LiMn_2O_4$ was confirmed using x-ray diffraction experiments performed on a x-ray diffractometer (Bruker D8 Discover, Germany) in a Bragg-Brentano geometry with Cu-K$_\propto$ radiation ($\lambda = 1.5406$ Å). The morphology of the as-received $V_2O_5$, as-prepared $LiV_3O_8$, and as-received $LiMn_2O_4$ was recorded using a field emission scanning electron microscope (Zeiss Gemini 500, Germany).

### 2.3. Electrochemical Characterization

The $V_2O_5$ electrode coatings were prepared over battery grade Al foil using an electrophoretic deposition technique reported by elsewhere.[41, 42] For the electrode coatings of $LiV_3O_8$ and $LiMn_2O_4$, conventional slurry based tape cast method was adopted. $LiMn_2O_4$ electrode coatings were prepared in a composition of active material: carbon black: binder = 7.5:1.5:1. $LiV_3O_8$ electrode coatings were prepared in a composition of active material: carbon black: binder = 3:1:1. Circular electrode discs of 15 mm diameter were punched out using a disc cutter (MSK-T06 MTI Corporation, USA). Electrodes with different active material loading, ranging from 1.2 – 6mg (in 15mm discs), were tested. Lithium ion half-cell configuration of CR2032 coin cells were fabricated using the prepared electrodes with lithium foil as counter and reference electrode. The electrolytes used in the tests were self-prepared inside an argon filled glove box with <0.5ppm for both $H_2O$ and $O_2$ (MBraun Labstar Pro). The various compositions tested are prepared according to the details mentioned in Table 1. The coin cells were assembled in an argon filled glove box (Mbraun, Germany), with <0.5ppm levels for both $O_2$ and $H_2O$.

Galvanostatic charge discharge studies were carried out in automated battery testers (BST8-MA, MTI Corporation and BTS4000-5V10mA, Neware) between 1.5V-3.8V (for $V_2O_5$), between 1.5V-4.0V (for $LiV_3O_8$), and between 3.0V-4.3V (for $LiMn_2O_4$) vs $Li^+/Li$ redox couple. The ionic conductivity of the electrolytes were measured using Autolab Microcell HC installed with TSC1600 closed electrochemical cell. The cell constant of the TSC1600 closed electrochemical cell was measured using 0.01D and 0.1D KCl solution at $25^oC$ (as per NIST standards), and calculated to be 22.622 $cm^{-1}$. EIS spectra measured for the electrolytes, recorded between 100kHz and 100Hz, were analyzed using ZSimpWin 3.21 program and Nova 2.1.5.[43]

**Table1:** Number of ions and solvent molecules present in the simulation box

| CODE | COMPOSITION | # $Li^+$ | # $TFSI^-$ | # DOL molecules | # DME molecules | # $V^{3+}$ ions (charge balanced by $OH^-$) |
|------|-------------|----------|------------|-----------------|-----------------|---------------------------------------------|
| EL1 | 1M LiTFSI in DOL-DME (1:1 v:v) | 602 | 602 | 4306 | 2900 | 100 (300) |
| EL2p5 | 2.5M LiTFSI in DOL-DME (1:1 v:v) | 1505 | 1505 | 4306 | 2900 | 100 (300) |

| EL4 | 4M LiTFSI in DOL-DME (1:1 v:v) | 2408 | 2408 | 4306 | 2900 | 100 (300) |
|------|------|------|------|------|------|------|
| EL5p5 | 5.5M LiTFSI in DOL-DME (1:1 v:v) | 3311 | 3311 | 4306 | 2900 | 100 (300) |
| EL7 | 7M LiTFSI in DOL-DME (1:1 v:v) | 4215 | 4215 | 4306 | 2900 | 100 (300) |

## 3. RESULTS AND DISCUSSION

### 3.1. *Physico-chemical property evaluation of different electrolyte systems under study*

The proposed molecular dynamics simulation model and force-field parameters can be validated by evaluating several macroscopic observables from the simulations, and comparing them against the experimentally obtained values. We begin this validation by comparing the calculated density, ionic conductivity and viscosity of the electrolytes (without the added vanadium ions), with the values experimentally measured by us (along with the values reported by other researchers).

Figure 1 shows the calculated density, ionic conductivity and viscosity of the different simulated electrolyte compositions. An increase in density is observed as the salt concentration increases from 1M to 7M. Since EL1 is a frequently studied electrolyte composition as a part of Li-S batteries, it is important that the physico-chemical properties calculated from our MD model matches well with the values reported in the literature. As shown in Figure 1(a), the calculated density for EL1 matches well with some of the experimental results reported in the literature with a deviation <2%.[44-47] The deviations between the measured and calculated densities for other compositions are also observed to be within 3%. This indicates that the force-field parameters used in the simulations should correctly predict the ion-ion and ion-solvent interactions. The densities calculated for pure DOL and pure DME are also observed to match well with the literature reported experimental values, as shown in Supporting Information. The self-diffusion coefficient of $Li^+$, $TFSI^-$, DOL, and DME for EL1, presented in Figure 1(b), is calculated from the intercept of a linear fit of MSD vs. time in log-log scale. The slope of the linear fit, which indicates the power dependency of MSD with time, is close to 1 for all the cases. This indicates that the self-diffusion of the ions and solvent exhibit a Fickian-type behavior. The calculated self-diffusion coefficients and the values reported in the literature are also observed to be of the same order of magnitude.[48-51] The large deviation in the experimentally reported values makes it difficult to correctly compare our calculated values with the experimental ones. Furthermore, non-polarizable force-fields are known to generate deviations of several orders of magnitudes, which are not observed in our case due to appropriate charge scaling.[52-56] For eg. Rajput et al. have also calculated self-diffusion coefficients of $Li^+$, $TFSI^-$, DOL, and DME, which are over 1 order of magnitude different from the experimentally obtained values.[49] The calculated ionic conductivity and shear viscosity of the different compositions of the electrolyte are shown in Figure 1(c). For the ionic conductivity, the Nernst-Einstein equation is not appropriate since it is derived for an infinitely dilute electrolyte system, wherein the ion-ion interactions are negligible. Therefore, a form of the ionic conductivity which accounts for the ion-ion interactions has been used to calculate the values for the concentrated electrolytes.[36] The ionic conductivity is observed to decrease as the concentration of LiTFSI in the electrolytes increases, which indicates that the strong ion-ion interactions hinder the movement of the ionic clusters in the concentrated electrolytes. This is also indicated by the reduced self-diffusion coefficient of the ions and the solvent molecules as the concentration of LiTFSI is increased (Supporting Information). As the concentration of LiTFSI is increased, a subdiffusive behavior is observed for the ions and the solvent molecules. The increase in the calculated shear viscosity also supports this reasoning. Suo et al. have reported a systematic experimental estimations of ionic conductivity and viscosity of electrolytes with varying concentration of LiTFSI (ranging from 1M-7M).[57] These values are plotted against our calculated and experimentally obtained results for comparison in Figures 1(c) and (d).While the ionic conductivity and shear viscosity follow the same trend as reported by Suo et al., a large deviation between the experimental and calculated values is observed for the shear viscosity. We believe that this discrepancy is due to the use of non-polarizable force-field, where our partial charge scaling factor should ideally be modified depending on the concentration of LiTFSI. As the concentration of LiTFSI in the system is increased, the effective screening of the ions from the solvent molecules is reduced due to the sheer number of the ions introduced into the system. Therefore, the partial charge scaling factor should be larger at higher concentrations. The calculated and the observed values are nearly equivalent for the compositions EL1-EL4. The deviations begin to appear when the concentration of LiTFSI is increased beyond 4M. Nevertheless, the calculated values do not differ significantly from the experimental ones. Therefore, the simulation model under study is suitable to gain a deeper understanding of the behavior of such concentrated electrolyte systems.

*3.2.Ion Dynamics and Transport Properties*

In this section, we study how the vanadium ions interact with the other ionic and molecular species present in the electrolytes. The vanadium ions can be introduced into the electrolytes in real systems from the dissolution of cathode materials during battery operation.[17-19] Such dissolution is known to be detrimental to battery performance. It is of interest to study how the transport properties of these dissolved vanadium ions are affected when the electrolytes are changed from a salt-in-solvent type (EL1) to a solvent-in-salt type (EL7). We begin our analysis by elucidating the local environment of the vanadium and lithium ions in the different electrolyte compositions using Radial Distribution Functions (RDF) and first solvation shell co-ordination numbers (CN). Judging from the partial charges present in the different ions and solvent entities, we expect the O atoms present in the $TFSI^-$, DOL, and DME to be strongly co-ordinated with the vanadium and lithium ions. Additionally, the hydroxide anion, which is added to make the system charge neutral, will also be co-coordinating with the vanadium and lithium ions. Figures 2(a)-(d) show the partial radial distribution functions for the $V^{3+}$-O and $Li^+$-O pairs in EL1 and EL7, with O being contributed from $TFSI^-$, DOL, and DME. The first solvation shell for $V^{3+}$ ions in EL1 are found to consist of $OH^-$, DOL, and DME, with a total coordination number of approximately 6. No $TFSI^-$ anions are present within the first solvation shell. The average coordination number with DME is calculated to be ~3, while it is calculated to be ~1 and ~2 with DOL and $OH^-$, respectively. On the other end, the first solvation shell for $V^{3+}$ ions in EL7 consist mostly of $TFSI^-$, DOL, and DME, with a total coordination number of approximately 6. The average coordination number with $TFSI^-$ is calculated to be ~2.1, while it is calculated to be ~2.2 and ~1.6 with DOL and DME, respectively. A similar trend is observed for the first solvation shell of $Li^+$ ions in EL1, with DME solvating the lithium-ions and no $TFSI^-$ present within the solvation shell. The first solvation shell of $Li^+$ for the EL7 case consist of $TFSI^-$, DOL, and DME. The difference in the number of $TFSI^-$ present within the first solvation shell of $V^{3+}$ ions will surely affect the way they are expected to behave in the two electrolyte systems. A snapshot of the first solvation shell of a vanadium ion in EL1 and EL7 is presented in Figure 3(a) and (b), respectively. The main coordinating centers from TFSI⁻, DOL, and DME to the vanadium ion turn out to be the oxygen species, matching well with the RDF data.

RDF measurements are generally performed in order to understand the static properties of the electrolyte. Study of the dynamic transport properties require suitable clustering of sets of molecules, and correlating their motion with each other. In order to study the dynamic transport properties of vanadium ions in the different electrolyte compositions, we first create suitable clusters with the vanadium ions sitting at the center. We choose center of mass of the molecules as the molecule identifier for this purpose, since center of mass is a suitable parameter to identify the molecules as a whole instead of mapping the positions of each of the atoms present within the molecules. From RDF measurements, it is expected that the cutoff radius for the first solvation shell should be around 5Å, with the second solvation shell thickness being roughly 3Å. Figures 4(a)-(e) show a 2D histogram of the number of vanadium ions containing an integral number of ions/molecules (TFSI⁻, DOL, and DME) within its second solvation shell (cutoff radius = 8Å) for all the electrolyte compositions. Very few TFSI⁻ are present within the second solvation shell of vanadium ions for electrolyte EL1. Calculations show that there is no TFSI⁻ present within the first solvation shell of vanadium in EL1. As the concentration of the salt increases in the electrolyte, the number of TFSI⁻ present within the second solvation shell also increases. At 7M concentration for EL7, most of the vanadium ions contain a maximum of 7 TFSI⁻ within the second solvation shell, which is not the case with the DME solvent molecules. The different solvation environment will also affect the solvation energy of the vanadium ions. A less negative free energy of solvation will thwart the dissolution of active material species into the electrolyte, which is known to cause cycleability issues in the batteries. Manganese and vanadium based compounds are especially notorious for undergoing active material dissolution into the electrolyte which fouls the separator, anode and the electrolyte. Therefore, it is of interest to calculate the free energy of solvation of the vanadium ions in the different electrolyte compositions, and record its changes as the electrolyte transitions from a salt-in-solvent type to solvent-in-salt type electrolyte. From the Born-Haber cycle, the enthalpy of dissolution can be written as the sum of lattice enthalpy and solvation enthalpy. The lattice enthalpy is material dependent, and electrolyte traditionally has no role in modifying its value. The solvation enthalpy is dependent on the type of electrolyte used. Therefore, a less negative solvation enthalpy will not favor the dissolution process. The free energy of solvation can be calculated by Free Energy Perturbation technique, where the interaction of the ion/species under consideration is reduced to zero in a graded manner. Figure 4(f) shows the free energy of solvation of the vanadium ion in different electrolyte compositions, along with the average number of TFSI⁻ present within the first solvation shell. It is observed that the free energy of solvation becomes less negative as the concentration of LiTFSI is increased in the electrolyte. This trend is positively correlated with the average number of TFSI⁻ present within the first solvation shell. This indicates that the dissolution process, from a thermodynamics point of view, should be arrested to a great extent in the solvent-in-salt type electrolytes.

While thermodynamics dictate that solvent-in-salt type superconcentrated electrolytes should be beneficial in obstructing the active material dissolution process during battery operation, the ionic clusters formed in these systems are strongly bound by electrostatic forces with their motions being highly correlated with each other. The dynamics of these clusters can be studied by constructing suitable residence-time based autocorrelation type functions. Since we are more concerned about how the ionic species in the electrolyte move under a normal battery operation within an internal electric field, we will focus only on the correlations between the motion of $V^{3+}$, $Li^+$ and $TFSI^-$ under the internal electric field of the battery. We define a residence-time type event by evaluating the number of active clusters (clusters having at least one $Li^+$/$TFSI^-$) which have undergone at least one deletion event, against time. The deletion event in this case is an event where a single ionic species moves out of the interaction radius of the $V^{3+}$ ion. Mathematically, this event can be represented as:

$$n_x(t') = \begin{cases} 0 \; ; \; r_{x-Y} > r_c \; \vee \; n_x(t) = 0 \; \ni t < t' \\ 1 \; ; \qquad\qquad\qquad otherwise \end{cases}, \; \forall \, x \in \{V^{3+}\}, Y \in \{\{Li^+\} \vee \{TFSI^-\}\}, \; r_{x-Y} <$$

$$r_i \; at \; t' = 0 \qquad\qquad\qquad\qquad\qquad\qquad (1)$$

The initial condition to define an active cluster for equation (1) is chosen such that the initial distance between the $V^{3+}$ and $Li^+$/$TFSI^-$ pair is less than a certain value $r_i$. We have kept $r_i$ equal to 8Å and 10Å for $TFSI^-$ and $Li^+$, respectively. The value of $r_c$ is kept at 10Å for both the cases. For each vanadium ion, the equation (1) will result in a step function. An average of multiple vanadium ions will transform this into a sigmoid curve, due to variable width of each of the step functions. This function is invariant under translations of time origin. Due to a comparison operation present with the earlier time steps to identify a deletion, a direct auto-correlation is not possible. However, averaging the measurements with different time origins will lead to a similar type of function being generated. Figures 5(a) and (b) show the  fraction of active clusters which did not undergo any deletion event against time, for $V^{3+}$-$Li^+$ and $V^{3+}$-$TFSI^-$ ion pairs, at different concentrations. It is observed that $t_{50}$, which represents the time where 50% of the clusters are yet to undergo a deletion event, is higher for the $V^{3+}$-$TFSI^-$ pairs as the concentration of the salt increases. On the contrary, $V^{3+}$-$Li^+$ clusters are short-lived, and have very low $t_{50}$ values as the concentration of the salt increases. This is surprising since the cations try to move along the direction of the electric field while the anions move in the opposite direction within a normal operation. Therefore, the motion of the $Li^+$ and $V^{3+}$ should be positively correlated and have lower deletion events occurring under normal circumstances. The opposite should be present for the $V^{3+}$-$TFSI^-$ pairs, where they should undergo more deletion events due to the motion of $TFSI^-$ and $V^{3+}$ being negatively correlated. For EL1, this behavior is noted with the $t_{50}^{V-Li} > t_{50}^{V-TFSI}$. The phenomenon is pictorially depicted for this case in Figure 5(c). The $TFSI^-$ ions, which are present within the second solvation shell, move out of the solvation shell in the direction opposite to that of the motion of the cluster under the presence of an electric field. The $Li^+$ ions move along with the cluster in the direction of the electric field, and hence undergo fewer deletions. In fact, about 40% of the clusters still haven't undergone any deletion event after 500ps of simulation run. As the concentration of the salt increases, more and more $TFSI^-$ ions are present within the solvation environment of the vanadium ions. Due to the increased coulombic interactions between $V^{3+}$ and $TFSI^-$, the motion of $V^{3+}$ ions becomes more and more restricted. Beyond a certain value of the concentration, an inversion in the behavior takes place wherein the $TFSI^-$ drags the $V^{3+}$ ion along with it. This leads to a negative transference for the vanadium ions since they are moving opposite to their concentration gradient. The phenomenon is pictorially depicted for this case in Figure 5(d). For EL7, the $t_{50}^{V-TFSI}$ is much larger than $t_{50}^{V-Li}$. If we refer to the data presented in Figure 4(e), we can observe that there are significantly high numbers of vanadium ion clusters which have over 3 $TFSI^-$ within the second solvation shell, making them negatively charged. Therefore, the cluster as a whole moves against the electric field direction all while dragging the vanadium ion with it. It is also observed that the $TFSI^-$ ions which are present in the first solvation shell of vanadium ions in EL7 rarely get detached and move out of the interaction volume of vanadium ions ($r_i = 5Å$) (Supporting Information).

These calculated results carry some serious consequences in the operation of a lithium-ion battery containing vanadium based cathode materials. From a thermodynamics point of view, the dissolution process of the active material should be retarded due to less negative free energy of solvation as one switches from salt-in-solvent type electrolytes to superconcentrated solvent-in-salt type electrolytes. In addition to this, the ion dynamics in the solvent-in-salt type electrolytes forces the dissolved ions to move against their concentration gradient, thus protecting the anode from fouling. This ultimately leads to better long-term cycleability of the batteries. These results can also be extended to other transition metal-ion cathode systems which are known to undergo active material dissolution process such as manganese based cathodes (LMO, LNMO etc). Since the above results are not cathode material specific, better long-term cycleability results should be achieved by changing the electrolytes to superconcentrated ones. A preliminary experimental validation of the hypothesis drawn from the present section is presented in the next section with $V_2O_5$, $LiV_3O_8$, and $LiMn_2O_4$ cathode materials.

### 3.3. Experimental Validation

As argued in the previous section, the free energy of solvation (not free energy of dissolution) is traditionally dependent only on the electrolyte. The particle morphology and phase has very little to no role in changing its value. Therefore, we test whether the cycleability of commercially available vanadium containing cathode materials is improved by transitioning into a superconcentrated electrolyte. We chose commercially available $V_2O_5$ to experimentally validate our model since its electrochemical activity is well studied, and it is known to undergo vanadium dissolution. We also synthesized $LiV_3O_8$ using reported recipes in the literature to validate our model since it is another well studied material reported to undergo vanadium dissolution. Figure 6(a) shows the cycleability curves for the $V_2O_5$ electrodes tested with EL1, EL4, and EL7 compositions at a specific current of 100mAg$^{-1}$. It is observed that the electrodes exhibit very poor cycleability with EL1, and undergoes a rapid capacity fading within 100 cycles. On the other hand, the electrodes exhibit excellent long-term cycleability with EL4 and EL7 compositions. While the initial specific capacity drops due to irreversible phase transitions are visible in all the cases, the specific capacity is observed to attain nearly constant values in the longer run for the concentrated electrolytes. Similar behavior is observed for the $LiV_3O_8$ material synthesized using protocols described in the literature, as shown in Figure 6(b). In the earlier part of this study, Mitra et al. demonstated that the cycleability of the Kazakhstanite phase ($Fe_5V_{15}O_{39}(OH)_9.9H_2O$) is also improved by transitioning into a superconcentrated electrolyte. [Unpublished work] Several other authors have also reported improved cycleability of vanadium containing cathode materials with superconcentrated electrolytes.[16, 22] Hence, the improvement in the cycleability of vanadium containing cathode materials through the use of superconcentrated electrolytes is consistent with the results predicted from Molecular Dynamics simulations. The conclusions drawn from Molecular Dynamics simulations can be extended to manganese containing cathode materials as well. Due to the lower partial charge of manganese (II) ions, which are known to be the dissolved form in lithium-ion electrolytes, the effect of increasing the salt concentration on impeding the movement of the ions towards the anode will be less significant. In spite of this, the improvement in the cycleability of manganese containing cathode materials is expected. We tested the cycleability of commercial $LiMn_2O_4$ with 1M $LiPF_6$ in EC:DMC (3:7 v:v) and EL4 electrolytes, and observe that the long-term cycleability is improved when using superconcentrated electrolytes (Figure 6(c)). In spite of several repetitions,

EL1 electrolyte could not be successfully used to construct LMO half-cells since corrosion of Al current collector occurs in the operational voltage range due to the presence of LiTFSI salt. However, the improvement is not significant and the higher salt concentrations do not participate in impeding the dissolution process completely. Using electrolytes with higher salt concentrations such as EL7 resulted in poor specific capacity of $LiMn_2O_4$, which is not suitable for practical applications (Supporting Information). We believe further modifications to the electrolyte compositions will be beneficial to improve the solubility limit approach for its successful implementation on the Manganese containing cathode materials.

## 4. CONCLUSIONS

In this manuscript, we show that the Molecular Dynamics simulations provide a strong justification for the solubility limit approach, and can be extended to vanadium containing cathode materials undergoing active material dissolution during operation. The modified partial charges and force-field parameters result in the calculated macroscopic properties of the electrolytes being consistent with the experimental results. It is observed that the local environments of the vanadium and lithium ions are different as the electrolyte is changed from a relatively dilute to a superconcentrated one. The anionic moieties are observed to populate the solvation sphere of the vanadium and lithium ions in superconcentrated electrolytes, resulting in negatively charged ionic clusters. The free energy of solvation is calculated to have a lower magnitude in the case of superconcentrated electrolytes, which provides a valid reasoning behind the hindered dissolution process. The negatively charged vanadium ion clusters have reduced ionic movement towards the anode under the presence of the internal electric field of the electrochemical cell, which creates a kinetic barrier towards the active material dissolution process. The theoretical framework is supported with an experimental validation by observing the cycleability of commercial $V_2O_5$, lab-synthesized $LiV_3O_8$, and commercial $LiMn_2O_4$ with different electrolyte compositions. The experimental cycleability plots reveal that a stable specific capacity with good long-term cycleability is observed for $V_2O_5$ and $LiV_3O_8$ when a superconcentrated electrolyte is employed instead of a relatively dilute electrolyte. The improvement in the long-term cycleability of $LiMn_2O_4$ is observed as well with the use of superconcentrated electrolytes. However, the use of superconcentrated electrolytes does not arrest the dissolution process completely for the case of $LiMn_2O_4$, since minor capacity fading is still observed during the long-term cycling of the $LiMn_2O_4$ electrodes.

**ASSOCIATED CONTENT**

**Supporting Information**. The supporting information file contains the calculated densities of the pure solvents; the calculated diffusion coefficients of the species in different electrolyte compositions; the residence time plots for $V^{3+}$- $TFSI^-$ with different cutoff radius. The supporting information file also contains the scanning electron micrographs and x-ray diffractograms of the commercial $V_2O_5$, $LiMn_2O_4$, and as-prepared $LiV_3O_8$, along with the charge discharge profile for commercial $LiMn_2O_4$ with EL7 electrolyte.

AUTHOR INFORMATION

**Author Contributions**

The manuscript was prepared through contributions of all authors (experimentation, analysis and writing). All authors have given approval to the final version of the manuscript.

**Funding Sources**

S.D. and S.B.M would like to acknowledge the funding from MHRD for project code UAY_IITKGP_001 (UAY Phase II), with approval order number F.No. 35-8/2017-TS.1, Dt: 05-12-2017, to carry out few of the parts of the work. S.D. would like to acknowledge the partial financial support from MHRD vide sanction F.16-59/2011-TEL, dated 30-09-2011 to carry out few parts of the work. S.B.M would like to acknowledge partial financial support obtained from SERB, DST vide sanction EMR/2016/007537, dated 16-03-2018 and SGCIR grant of IIT Kharagpur vide approval No. IIT/SRIC/MS/MPV_ICG_2017_SGCIR/2018-19/080 dated 15-07-2018 to carry out few parts of the work. A.M. would like to acknowledge the funding received to carry out the research under the Prime Minister Research Fellowship (PMRF) scheme.

**ACKNOWLEDGMENT**

A.M would like to acknowledge MHRD, Govt. of India for the Prime Minister Research Fellowship. All the authors would like to acknowledge the use of the Supercomputing facility of IIT Kharagpur established under National Supercomputing Mission (NSM), Government of India and supported by Centre for Development of Advanced Computing (CDAC), Pune, to carry out the Molecular Dynamics simulations. The authors would like to acknowledge the FESEM facility sponsored by DST-FIST at Materials Science Centre, Indian Institute of Technology Kharagpur for scanning electron microscopy experiments. The authors would like to acknowledge the FESEM facility sponsored by DST-FIST at Department of Metallurgical and Materials Engineering, Indian Institute of Technology Kharagpur for scanning electron microscopy and energy dispersive x-ray spectroscopy experiments.

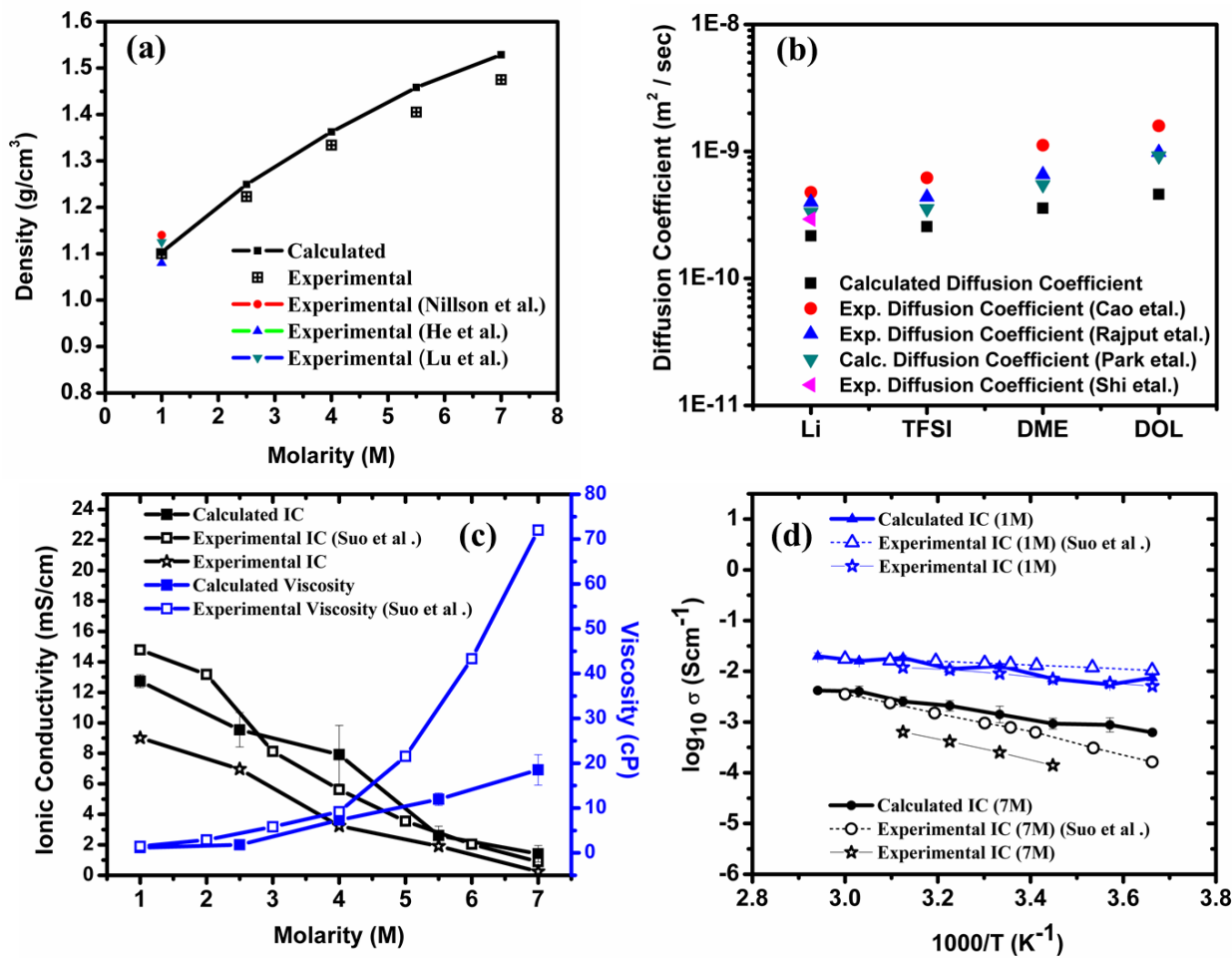

**Figure 1: (a)** The calculated and experimental densities of the various electrolyte compositions which are undertaken for the simulation studies.[44, 45, 47] **(b)** The calculated diffusion coefficients of the various ionic and solvent species in EL1 composition, along with their comparison with reported values in the literature.[48-51] **(c)** The calculated ionic conductivity and viscosity of the various electrolyte compositions tested for the simulation studies, along with a comparison with the experimental values.[57] **(d)** The temperature-dependent ionic

conductivity values for the EL1 (relatively dilute) and EL7 (superconcentrated) electrolytes [experimental and calculated]. [57]

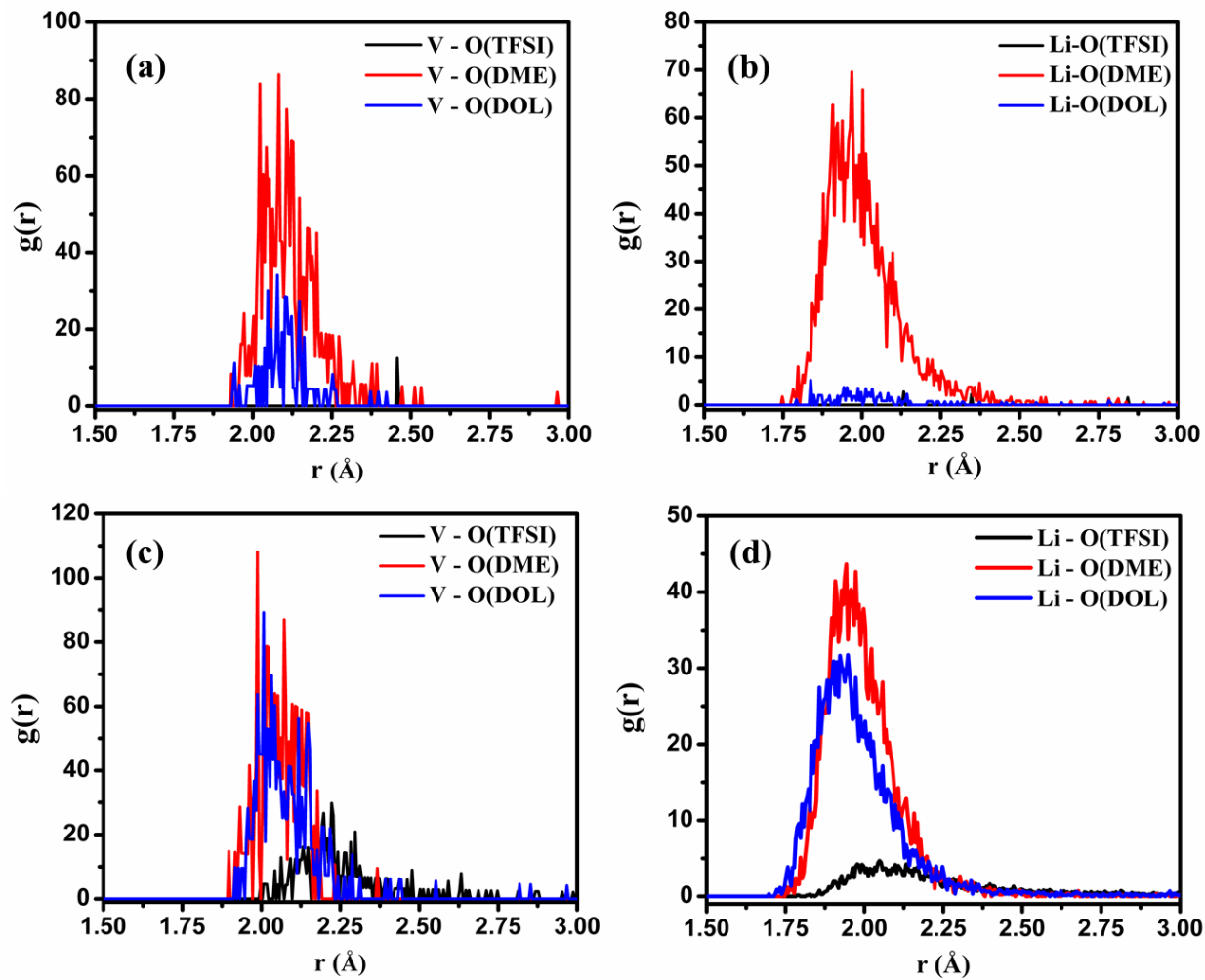

**Figure 2:** Extracted Radial Distribution Functions (RDF) for vanadium and lithium ions with a 3Å radius cutoff to show their co-ordination with the oxygen atoms from various species (TFSI⁻, DOL, and DME) in **(a)** EL1, and **(b)** EL7 compositions.

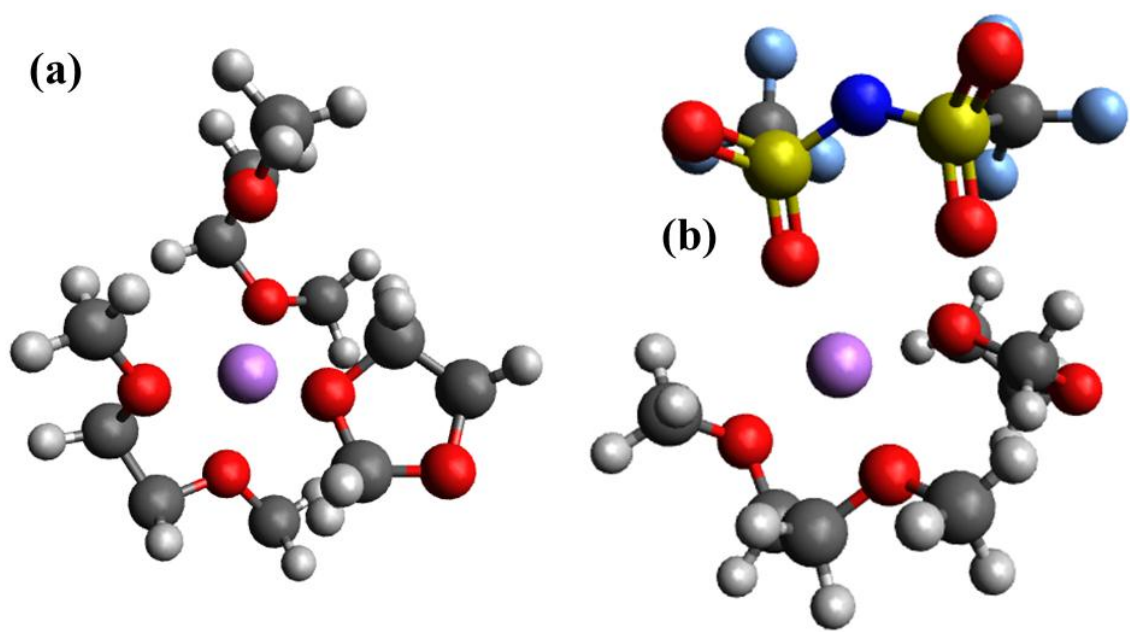

**Figure 3:** Snapshot of the co-ordination environment of the vanadium ions in first solvation shell in **(a)** EL1, and **(b)** EL7 compositions.

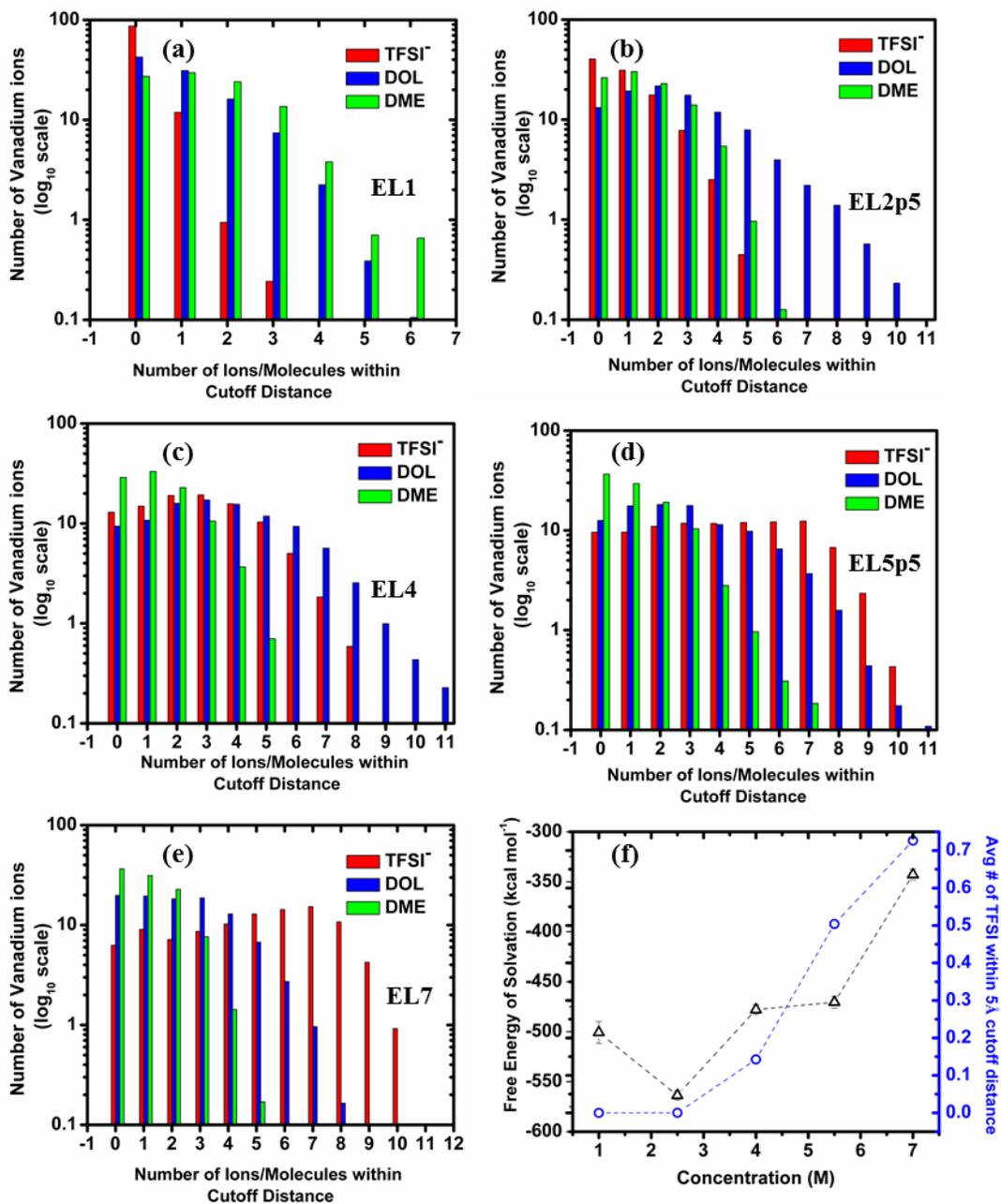

**Figure 4:** Histogram for the number of vanadium ions with a specific number of ions/molecules (TFSI⁻/DOL/DME) within a cutoff radius of 8Å (second solvation shell) for the electrolyte compositions **(a)**EL1, **(b)** EL2p5, **(c)** EL4, **(d)** EL5p5, and **(e)** EL7. **(f)** The free energy of solvation of vanadium ions calculated from the FEP technique for different electrolyte

compositions, along with a correlation with the average number of TFSI$^-$ within first solvation shell.

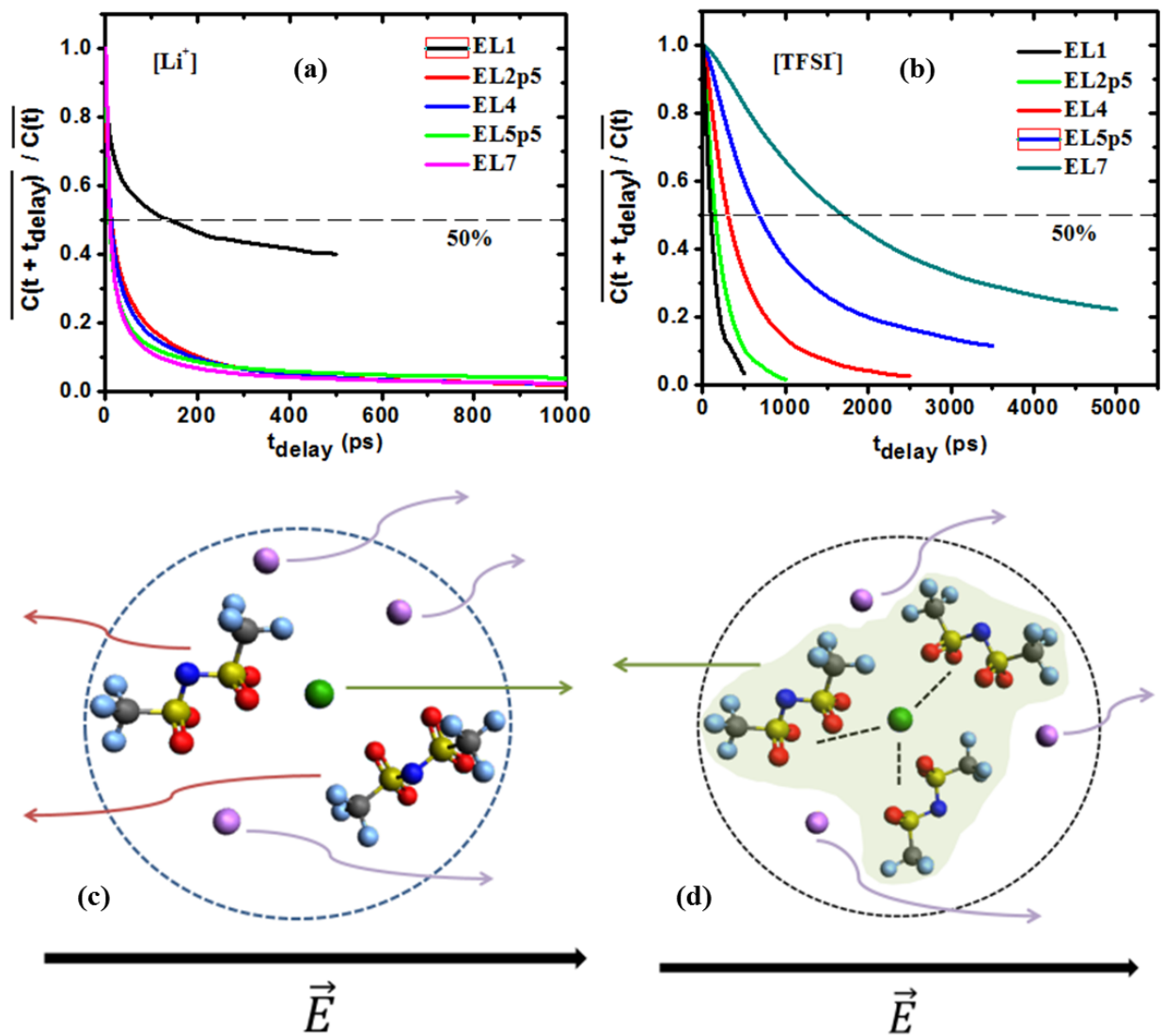

**Figure 5:** Fraction of active clusters which did not undergo any deletion event against time according to the equation (1), for **(a)** V$^{3+}$-Li$^+$ and **(b)** V$^{3+}$-TFSI$^-$ ion pairs, at different concentrations. (c) Pictorial representation of the ion movement process in the electrolyte composition EL1. (d) Pictorial representation of the ion movement process in the electrolyte composition EL7.

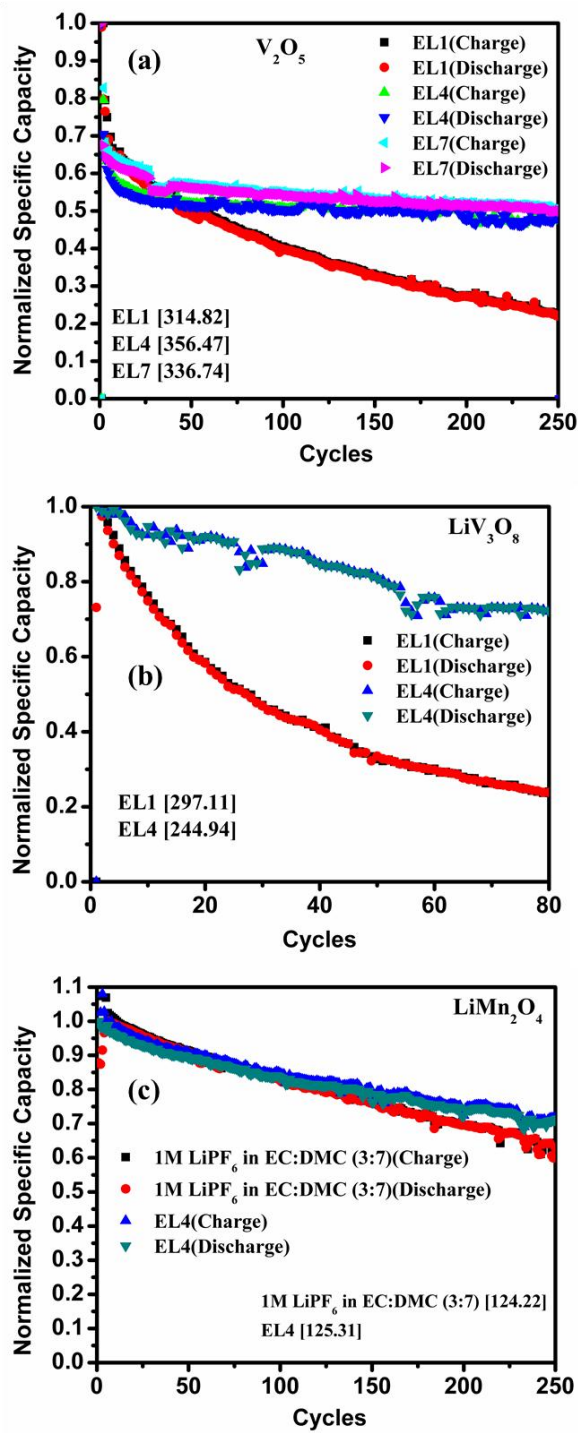

**Figure 6:** The cycleability plots for **(a)** commercial $V_2O_5$, **(b)** as-synthesized $LiV_3O_8$, and **(c)** commercial $LiMn_2O_4$ electrodes with different electrolyte compositions. The normalization factors are presented in the square brackets in the figures.